\documentclass[a4paper,11pt]{article}
\pdfoutput=1 

\usepackage{jcappub} 

\usepackage[T1]{fontenc} 

\usepackage{natbib}

\usepackage{multirow}
\graphicspath{{./images/}}

\def\ref@jnl#1{{\jnl@style#1}}

\def\aj{\ref@jnl{AJ}}                   
\def\actaa{\ref@jnl{Acta Astron.}}      
\def\araa{\ref@jnl{ARA\&A}}             
\def\apj{\ref@jnl{ApJ}}                 
\def\apjl{\ref@jnl{ApJ}}                
\def\apjs{\ref@jnl{ApJS}}               
\def\ao{\ref@jnl{Appl.~Opt.}}           
\def\apss{\ref@jnl{Ap\&SS}}             
\def\aap{\ref@jnl{A\&A}}                
\def\aapr{\ref@jnl{A\&A~Rev.}}          
\def\aaps{\ref@jnl{A\&AS}}              
\def\azh{\ref@jnl{AZh}}                 
\def\baas{\ref@jnl{BAAS}}               
\def\bac{\ref@jnl{Bull. astr. Inst. Czechosl.}}
\def\caa{\ref@jnl{Chinese Astron. Astrophys.}}
\def\cjaa{\ref@jnl{Chinese J. Astron. Astrophys.}}
\def\icarus{\ref@jnl{Icarus}}           
\def\jcap{\ref@jnl{J. Cosmology Astropart. Phys.}}
\def\jrasc{\ref@jnl{JRASC}}             
\def\memras{\ref@jnl{MmRAS}}            
\def\mnras{\ref@jnl{MNRAS}}             
\def\na{\ref@jnl{New A}}                
\def\nar{\ref@jnl{New A Rev.}}          
\def\pra{\ref@jnl{Phys.~Rev.~A}}        
\def\prb{\ref@jnl{Phys.~Rev.~B}}        
\def\prc{\ref@jnl{Phys.~Rev.~C}}        
\def\prd{\ref@jnl{Phys.~Rev.~D}}        
\def\pre{\ref@jnl{Phys.~Rev.~E}}        
\def\prl{\ref@jnl{Phys.~Rev.~Lett.}}    
\def\pasa{\ref@jnl{PASA}}               
\def\pasp{\ref@jnl{PASP}}               
\def\pasj{\ref@jnl{PASJ}}               
\def\rmxaa{\ref@jnl{Rev. Mexicana Astron. Astrofis.}}%
\def\qjras{\ref@jnl{QJRAS}}             
\def\skytel{\ref@jnl{S\&T}}             
\def\solphys{\ref@jnl{Sol.~Phys.}}      
\def\sovast{\ref@jnl{Soviet~Ast.}}      
\def\ssr{\ref@jnl{Space~Sci.~Rev.}}     
\def\zap{\ref@jnl{ZAp}}                 
\def\nat{\ref@jnl{Nature}}              
\def\iaucirc{\ref@jnl{IAU~Circ.}}       
\def\aplett{\ref@jnl{Astrophys.~Lett.}} 
\def\apspr{\ref@jnl{Astrophys.~Space~Phys.~Res.}}
\def\bain{\ref@jnl{Bull.~Astron.~Inst.~Netherlands}} 
\def\fcp{\ref@jnl{Fund.~Cosmic~Phys.}}  
\def\gca{\ref@jnl{Geochim.~Cosmochim.~Acta}}   
\def\grl{\ref@jnl{Geophys.~Res.~Lett.}} 
\def\jcp{\ref@jnl{J.~Chem.~Phys.}}      
\def\jgr{\ref@jnl{J.~Geophys.~Res.}}    
\def\jqsrt{\ref@jnl{J.~Quant.~Spec.~Radiat.~Transf.}}
\def\memsai{\ref@jnl{Mem.~Soc.~Astron.~Italiana}}
\def\nphysa{\ref@jnl{Nucl.~Phys.~A}}   
\def\physrep{\ref@jnl{Phys.~Rep.}}   
\def\physscr{\ref@jnl{Phys.~Scr}}   
\def\planss{\ref@jnl{Planet.~Space~Sci.}}   
\def\procspie{\ref@jnl{Proc.~SPIE}}   

\title{Reconstruction of the Dark Energy equation of state from latest data: the impact of theoretical priors.}

\author[a,b,1]{Francesca Gerardi,}
\author[b]{Matteo Martinelli,\note{Corresponding author.}}
\author[b]{Alessandra Silvestri}

\affiliation[a]{Dipartimento di Fisica ed Astronomia ``Galileo Galilei'', Universit\`a di Padova,\\ Vicolo Osservatorio 3, I-35122 Padova, Italy}
\affiliation[b]{Institute Lorentz, Leiden University, PO Box 9506, Leiden 2300 RA, The Netherlands}

\emailAdd{francesca.gerardi@studenti.unipd.it}
\emailAdd{martinelli@lorentz.leidenuniv.nl}
\emailAdd{silvestri@lorentz.leidenuniv.nl}

\abstract{We reconstruct the Equation of State of Dark Energy (EoS) from current data using a non-parametric approach where, rather than assuming a specific time evolution of this function, we bin it in time. We treat the transition between the bins with two different methods, i.e.  a smoothed step function and a Gaussian Process reconstruction, investigating whether or not the two approaches lead to compatible results. Additionally, we include in the reconstruction procedure a correlation between the values of the EoS at different times in the form of a theoretical prior that takes into account a set of viability and stability requirements that one can impose on models alternative to $\Lambda$CDM. In such case, we necessarily specialize to broad, but specific classes of alternative models, i.e.  Quintessence and Horndeski gravity. We use data coming from CMB, Supernovae and BAO surveys. We find an overall agreement between the different reconstruction methods used; with both approaches, we find a time dependence of the mean of the reconstruction, with different trends depending on the class of model studied. The constant EoS predicted by the $\Lambda$CDM model falls anyway within the $1\sigma$ bounds of our analysis.}

\begin{document}
\maketitle
\flushbottom

\section{Introduction}
\label{sec:intro}

Since the discovery of the late time cosmic acceleration~\citep{Riess1998, Perlmutter1998}, the physical mechanism underlying this accelerated phase has been an elusive one and still poses an open question for Cosmology. The simplest candidate that could drive this acceleration is a Cosmological Constant $\Lambda$, as in the $\Lambda$CDM model. Despite its success in describing cosmological observations~\citep{Planck2018}, the $\Lambda$CDM model poses theoretical questions that still prompt the investigation of the nature of $\Lambda$, i.e. the value of this constant and the special moment of the Universe lifetime in which this starts to dominate over the other components~\citep{Weinberg,Sahni:1999gb,Burgess} .

Furthermore, recent observations have reached a precision which started to highlight tensions between the measurements of $\Lambda$CDM parameters as inferred from different probes. The most striking one, appears in the values for the Hubble constant $H_0$ found with local measurements \citep{Riess_H0} and with Cosmic Microwave Background (CMB) observations \citep{Planck2018}. While the former directly measure the current rate of expansion, the latter rely on the assumption of a specific cosmological model ($\Lambda$CDM) to extrapolate high redshift measurements to present time. A similar tension, although less statistically significant, is also found when measuring the growth of cosmological perturbations from a high or low redshift point of view, relying again on CMB \citep{Planck2018} versus galaxy surveys \citep{CFHTLenS,KiDS-450,DES1}. Even though it is not excluded that these tensions could be driven by systematic errors, it could also be that they are due to the assumption of the $\Lambda$CDM model in the analysis of high redshift data. 

Many alternatives to the $\Lambda$CDM  have been put to test against observations (see e.g. \citep{Davis:2007na,Amendola:2007rr,Chiba:2012cb,Cataneo:2014kaa,Bonilla:2017ygx,Wen:2017aaa}) sometimes allowing to rule out specific theories \citep{Fairbairn:2005ue,Maartens:2006yt,Rydbeck:2007gy,Song:2006jk} , but more often leading only to constraints of their parameter space around the $\Lambda$CDM limit. Given the  significant amount of available models and the difficulty in testing all of them against the data, a model independent approach is desirable and the first efforts in developing such an approach date back to almost twenty years ago, right after the discovery of the late acceleration phenomenon \citep{Huterer:1998qv, Chiba:2000im, Saini:1999ba, Huterer:2000mj}. A typical example is the so-called Chevallier-Polarski-Linder (CPL) parameterization \citep{Chevallier_wde, Linder_wde}, a simple two parameters description of the time-dependence of the equation of state for the component that dominates at late times and sources the cosmic acceleration. We will generically refer to this component as Dark Energy (DE), and refer to its equation of state as $w_{\rm DE}$ (EoS). This could be an actual additional fluid contributing to the energy-momentum tensor, or, alternatively, result from modifications of gravity (MG). In either cases, generally $w_{\rm DE}$ is expected to be non-zero. The CPL parameterization has been used extensively in the literature to constrain the time evolution of DE (see e.g. \citep{Planck2015_DE_MG,Planck2018,CMB_BAO_SNe,Scherrer:2015tra}), and also to quantify the ability of future experiments to shed light on the nature of DE \citep{Euclid_greco}.
Even though this approach allows to investigate the DE problem without restricting to a specific theory, it still relies on assumptions on the time evolution of $w_{\rm DE}$. An alternative approach is to reconstruct $w_{DE}$ constraining its value at different times. This so-called ``non-parametric'' approach  has been often used in literature, and used to constrain the expansion history with data such as Type Ia SNe \citep{Shafieloo:2005nd, NPA_2010, NPA_2012,   Porqueres:2016kfv} along with BAO \citep{Sahni:2006pa, NPA_2010_2, Said, LHuillier:2016mtc, NPA_2018, Shafieloo:2018gin}, but the amount of free parameters needed to reconstruct the EoS has often hindered its usefulness.

In this paper, we take one step further and combine theoretical priors into the non-parametric reconstruction of $w_{\rm DE}$, based on general requirements of physical viability \citep{Articolo_priors_Ale}. These  are derived for broad, but specific classes of theories. Hence, even though they are mild,  including them in the reconstruction slightly limits the model independence. More importantly, the priors allow to introduce a, theoretically informed correlation between the values of $w_{\rm DE}$ at different times, aiding the reconstruction. Such an approach also guarantees that the reconstructed function will eventually correspond to a theoretically viable model, while  it can be complicated to map a completely model independent reconstruction of the EoS to the one produced by any well behaved theory \citep{Peirone:2017lgi}.

The paper is organized as follows. In Section \ref{sec:npr} we outline the reconstruction techniques we exploited to obtain a model independent $w_{\rm DE}$ from binned values constrained through observational data. In Section \ref{sec:data_analysis_method} we describe the general data analysis methodology employed, we describe the theoretical information coming from viability requirements for two classes of alternatives to $\Lambda$CDM (Quintessence and Horndeski), and we detail how this information is included in our analysis. We then present the results of this reconstruction in Section \ref{sec:results}, before drawing our conclusions in Section \ref{sec:concl}.

\section{Non parametric reconstruction}\label{sec:npr}
We start discretizing the Dark Energy EoS,  $w_{\rm DE}$, into several binned values, with the value of $w_{\rm DE}$ in each bin being  a free parameter of our analysis. The evolution of $w_{\rm DE}$ in time can be equivalently expressed in terms of the redshift $z$ or of the scale factor $a$, which are connected to time through the Friedmann equations.
In this paper we choose to discretize $w_{\rm DE}$ into bins equally spaced in the scale factor $a$ (with  $w_{\rm DE}(a_{i})=w_i$ at the center of the $i$-th bin), in the interval $[a_{\rm min}, 1]$. The choice of binning the EoS in scale factor rather than in redshift, as well as the specific choice of the binning strategy, is due to the analysis method that we will detail further in this paper, in particular to the use of theoretical viability priors (see Section \ref{subsec:corr_prior}). Indeed, accounting for theoretical considerations on the general behavior of $w_{\rm DE}(a)$ would set conditions on the reconstruction and on the correlation between the binned values of the function, e.g. motivating a correlation length $\xi$ between the $w_i$ and, consequently, the setting of  the binning strategy \citep{Crittenden_2009}. We will describe the binning strategy more explicitly case by case in Section~\ref{subsec:corr_prior} .

Cosmological observables generally do not depend directly on the values of $w_{\rm DE}$ at each redshift, but rather on the evolution in time of this function between the observer and the measured source. Therefore, we need to join the binned values into a function of time that can be, e.g.,  integrated. Let us use the luminosity distance  $d_L(a)$ inferred from observations of standard candles to illustrate this point. It can be expressed as 
\begin{equation}
d_L(a) = \frac{c}{aH_0}\int_a^1{\frac{da'}{E(a')}}
\end{equation}

with 
\begin{equation}
E(a) \approx \sqrt{\Omega_ma^{-3}+\Omega_{\rm DE}\exp{\left[\int_a^1{\frac{3[1+w_{\rm DE}(a)]}{a}da'}\right]}}.
\end{equation}

Clearly, if we are to fit a binned EoS to SNe data, we need to ensure a well defined evolution for $w_{\rm DE}$ by specifying how the function will behave within a certain bin and in moving from one to the other. Several reconstruction strategies have been used in past literature ranging from simple or smoothed step functions to more complex statistical tools (see e.g. \cite{NPA_2012, Said, LHuillier:2016mtc, Porqueres:2016kfv}); in our analysis we interpolate the binned $w_{i}$ values via two alternative techniques: a smoothed step function and Gaussian Process (GP), described in detail below. Indeed, once the binning properties for a specific case are given, we develop two different reconstructions, corresponding to the two different methods, and compare the results obtained.

\subsection{Smoothed step function}
\begin{figure}[h!]
\begin{center}
\includegraphics[scale=0.3]{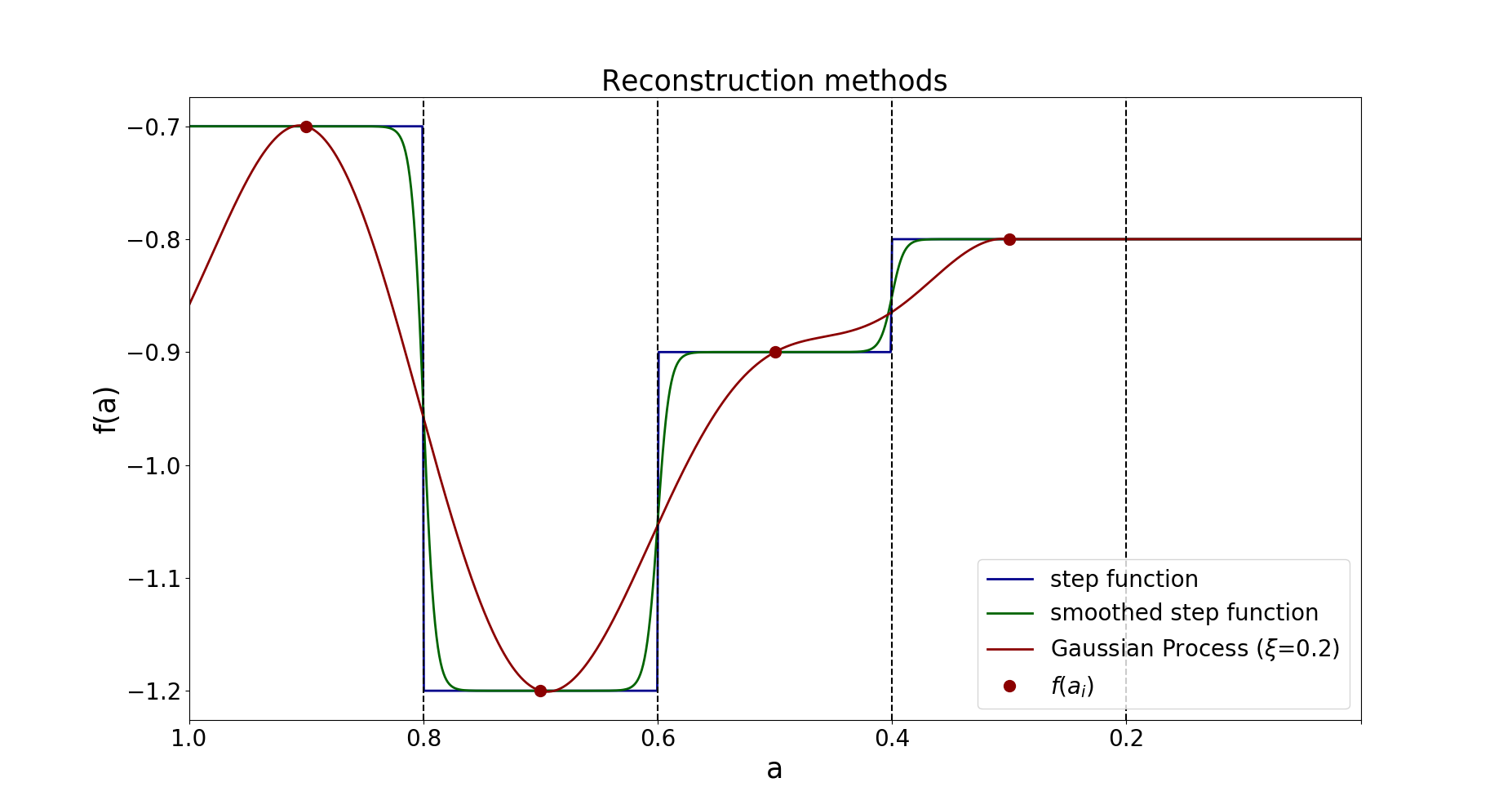}
\caption{Reconstructed $w_{\rm DE}$ as a function of the scale factor, with three different methods: step function, smoothed step function and Gaussian Process; the red dots are the chosen $w(a_{i})$. The values of $w_{\rm DE}$ at $a<a_{\rm min}$ are fixed to $w_{\rm DE}(a_{min})$.}\label{fig:camb_binning}
\end{center}
\end{figure}

The simplest choice to interpolate the binned values of $w_{\rm DE}$ is to use a step function. Defining $w_{\rm DE}(a_{i})=w_i$ as the values of the EoS at the center of each bin, 
\begin{equation}\label{eq:sharpbin}
w_{\rm DE}(a)=w_1+\sum_{i=1}^{N-1}\left(w_{i+1}-w_{1}\right)\left[\theta_H(a-a_i)-\theta_H(a-a_{i+1})\right],
\end{equation}
where $\theta_H$ is the Heaviside function and $N$ is the number of bins.
With this choice the EoS is constant within each bin and it has a sharp transition in its value when moving from one bin to the next (see blue line in Figure \ref{fig:camb_binning}).

Such a reconstruction can in principle cause numerical issues due to the fact that the function and its derivative are not well defined at the boundaries between bins. For this reason, we adopt a smoothed step function (see e.g. \citep{MATTEOMG})

\begin{equation}\label{smooth_rec}
w_{\rm DE}(a) = w_1 + \sum_{i=1}^{N-1} \dfrac{w_{i+1} - w_{i}}{2} \bigg\{ 1+\tanh \left[ s \left( \dfrac{a-a_{i+1}}{a_{i+1}-a_{i}} \right)  \right]   \bigg\}
\end{equation}
where $N$ is the number of bins, $s$ is a smoothing factor used to control the slope of the transition from one bin to the adjacent ones. In this work we set $s=10$\footnote{The choice of this value is chosen by trial and error in order to find a smoothing parameter producing a smooth transition between neighbouring bins, but still preserving a constant value within the bins themselves. The value used here is the same used in previous reconstructions \cite{MATTEOMG}}; the resulting reconstruction can be seen in Figure \ref{fig:camb_binning} (green line).

\subsection{Gaussian Process}
In the smoothed step approach discussed above,  the reconstructed function is assumed to be constant within each redshift bin; this can in principle bias the results obtained if the choice of the bin intervals is not done properly. To overcome such a problem, one can rely on more sophisticated techniques, e.g. Gaussian Process.
A GP is defined as a collection of random variables , any finite number of which have a joint Gaussian distribution \citep{MIT}. Considering the training points  $\vec{a} = (a_{1}~a_{2}~a_{3})$, we can always think of the function $w_{\rm DE}(a)$ 
evaluated at these points as a vector and, at each point $a_{i}$, $w_i$ is a Gaussian random variable with mean $\mu(a_{i})$ and variance $\sigma_{a, i}^{2}$. The entire vector will then be modeled with a multivariate Gaussian distribution
\begin{equation}
\vec{w}_{\rm DE}\\
=
\begin{bmatrix}
w_{1}\\
w_{2}\\
w_{3}
\end{bmatrix}
\sim
\mathcal{N}(\vec{\mu},C)\\
=
\mathcal{N} \bigg( \vec{\mu},
\begin{bmatrix}
C_{11} & C_{12} & C_{13}\\
C_{21} & C_{22} & C_{23}\\
C_{31} & C_{33} & C_{33}
\end{bmatrix} \bigg)
\end{equation}
where $\mathcal{N}$ stands for Normal distribution and  $C$ is the covariance matrix, with $C_{ij}=C(w_i,w_j)$ denoting the correlation between two points $a_{i}$ and $a_{j}$. We expect that the closer $a_{i}$ and $a_{j}$ are, the more the corresponding values of the EoS,  $w_{i}$ and $w_{j}$,  will be correlated.
In order to interpolate the training points we use the python package \texttt{sklearn}, which provides several choices for the covariance $C$, and we work here with the Radial Basis Function (RBF) :
\begin{equation}\label{GP_covmat}
C(a,a') = e^{-\frac{\vert a-a' \vert ^{2}}{2 \xi^{2}}}
\end{equation}
where $\xi$ is the correlation length, such that
\begin{equation}\label{cov}
C(a,a') =
\bigg \{
\begin{array}{ll}
 0 & ~~\vert a-a' \vert \gg \xi \\
 1 & ~~a = a' \\
\end{array}
\end{equation}
Such a choice provides a Kernel for the GP which is both stationary, because it is a function of $a - a'$, and isotropic, since it is a function of its module $\vert a - a' \vert$. The reason why we chose this Kernel is that it is fully characterized by only one parameter, the correlation lenght $\xi$. Of course this results in a lower flexibility of our reconstruction, but it does not introduce new degeneracies associated to a high number of hyperparameters. Moreover, this kernel is infinitely differentiable, leading to the process being infinitely mean-square differentiable \citep{MIT}. In this paper, we do not explore the dependence of our results on the choice of the GP Kernel, but rather leave this investigation for our ongoing work which extends to the reconstruction of the functions relevant for large scale structure.

Given a set of training points $(a_{i}, w_i)$, we want to obtain the value of the function at a point $a_{*}$, defined as $w_*$. Since we expect the function to be smooth, for a small variation of the $a$ variable we do not expect the function at that point to differ much from its values in the adjacent points. 

Assuming that $w_*$ will be Gaussian distributed as well, i.e. $w_* \sim \mathcal{N}(\mu_{*},C(a_{*},a_{*}))$, where $C(a_{*},a_{*})$ is the self-covariance, the joint distribution will assume the form of 
\begin{equation}
\begin{bmatrix}
\vec{w}_{\rm DE}\\
w_{*}\\
\end{bmatrix}
\sim
\mathcal{N} \bigg( 
\begin{bmatrix}
\vec{\mu}\\
\mu_{*}\\
\end{bmatrix},
\begin{bmatrix}
C(\vec{a},\vec{{a}}') & C(\vec{a},a_{*}) \\
C(a_{*},\vec{a})     & C(a_{*},a_{*}) \\
\end{bmatrix} \bigg)
\end{equation}

After specifying the mean and correlation functions, considering now all the fitting points, the GP is defined as
\begin{equation}
w_{\rm DE}(\vec{a}) \sim GP (\mu(\vec{a}),C(\vec{a},\vec{a}'))
\end{equation}

This reconstruction method does not require only the binned points $a_{i}$ and $w_{i}(a)$ as an input, but also specific choices for the mean and the correlation length of the Gaussian distribution.
Usually, GPs are used on observational data where the training points are fixed (provided by the data), and therefore one can obtain $\mu$ and $\xi$ as the values providing the best fit of the reconstructed function to the observations (see e.g. \citep{NPA_2012,Haridasu:2018gqm}). Here instead, while the $a_i$ are fixed, the $w_i$ values are free parameters of our analysis. Therefore we need to obtain different GP reconstructions for each sampled set of parameters $w_i$. In order to achieve this, we choose not to vary the hyperparameters of the Gaussian process, $\mu$ and $\xi$. Indeed we fix them to set values, where $\xi$ is directly obtained from the theoretical priors, imposed by the requirement that the reconstructed $w_{\rm DE}(a)$ can be eventually linked to a physically viable theoretical model. In Section \ref{subsec:corr_prior} we will discuss in detail how specifying to classes of alternative models provides theoretical information on the correlation length, and we will provide the values used in the different cases.

This theoretical prior we will make use of will not provide however any information on the mean of the GP $\mu$. This means that a correct analysis would need to use $\mu$ as a free parameter. We point out however, that given our choice of datasets, with SN and BAO significantly helping CMB to constrain the redshift trend of $w_{\rm DE}$, we do not expect this function to be extremely different from the $\Lambda$CDM limit. Moreover, with a sensible choice of the bins used to reconstruct the function, we expect the choice of $\mu$ to not impact significantly the results: the GP will move the function towards the mean whenever the reconstruction is done at a scale factor that is more than a correlation length away from the training points. This can be an issue when using real data as training points, but in our case we decide the bin positions a priori, and our binning strategy is able to limit the impact of fixing the $\mu$ parameter.

As we will obtain our results using GP, but also the smoothed step function as an alternative reconstruction method, we will be able to verify these assumptions; the second method is not affected by the choice of the GP hyperparameters, therefore, when comparing the results, a significant discrepancy might imply that the choice of $\mu$ is affecting the reconstruction significantly. As such a discrepancy is not found (see Section \ref{sec:results}), we don't investigate further the dependence of the GP reconstruction on these parameters, leaving it for a future dedicated work.

\section{Data and analysis method} \label{sec:data_analysis_method}
Given the binned $w_{\rm DE}$, a reconstruction method and the standard set of cosmological parameters,  we can make predictions for the desired cosmological observables and constrain our parameter space against data. To this extent, we use the public code {\tt CAMB} \citep{CAMB1,CAMB2}, modifying it in such a way that we can provide as an input  the binned values of $w_{\rm DE}$ and the chosen reconstruction method. We then sample the entire parameter space with the public Monte Carlo Markov Chain (MCMC) code {\tt CosmoMC} \citep{Lewis_Sarah}. For the  cosmological parameters,  we choose those of the minimal flat $\Lambda$CDM cosmology: the baryon and cold dark matter densities at present day, $\Omega_b h^2$ and $\Omega_ch^2$; the optical depth, $\tau$; the primordial power spectrum amplitude and tilt, $A_s$ and $n_s$, and the Hubble parameter $H_0$. Alongside the standard cosmological parameters we also sample the binned values of the EoS $w_{i}$. The prior range and bins positions in scale factors will depend on the specific cases investigated and will be specified in the following section; in general however, we assume that the EoS stays constant outside the binned interval, i.e. $w(a>a_1)=w_1$ and $w(a<a_N)=w_N$. We use flat priors on all the parameters.

We use the following datasets: the JLA ('Joint Light-curve Analysis') dataset \citep{Betoule:2014frx}, that unifies Type Ia SNe observations of SDSS-II (Sloan Digital Sky Survey) \citep{SDSS-II} and SNLS (Supernova Legacy Survey) \citep{SNLS} collaborations, for a total of 740 Type Ia SNe up to redshift $z \sim 1$; the 6dFGS (6dF Galaxy Survey) \citep{BAO_6dF} and SDSS Data Release 7 \citep{BAO_MGS} for BAO and Planck 2015 data \citep{Planck2015} for CMB. We will specify in the following subsection the  prior ranges for $w_i$, as well as the different combinations of the data listed above, depending on the different classes of theories under consideration.

\subsection{Theoretical correlation prior}
\label{subsec:corr_prior}

The methods presented in Section \ref{sec:npr} allow to reconstruct the Dark Energy EoS as a function of redshift starting from its binned values; however the $w_i$ in each bin can in principle assume any value, with no relation to the values of the other bins. This does not take into account that viable theoretical models do not generally allow for extreme oscillations of this function \citep{Crittenden_2012}, and therefore such a reconstruction can lead to a $w_{\rm DE}(a)$ that cannot be linked to any physically viable  theoretical model. As already anticipated, we overcome this problem including a theoretically informed prior which account for the fact that a given $w_i$ will be correlated with the values assumed by the function in the adjacent bins. Following the approach of \citep{Crittenden_2009} we impose a correlation prior, which rescales the Likelihood function ($\mathcal{L}$) as
\begin{equation}
-\ln{\mathcal{L}}=\chi^2=\chi^2_{\rm data}+\chi^2_{\rm prior},
\end{equation} 
where we assumed a Gaussian distribution $\mathcal{L}\propto\exp{[-\chi^2/2]}$. We define this prior as
\begin{equation}
\chi^{2}_{\rm prior} = (\mathbf{w} - \mathbf{\bar{w}})^{T} \mathcal{C}^{-1} (\mathbf{w} - \mathbf{\bar{w}}),
\end{equation}
where $\mathbf{w}=\{w_1,...,w_N\}$ is the vector containing the binned values of $w_{\rm DE}(a)$, $\mathcal{C}$ is a covariance matrix and $\mathbf{\bar{w}}$ is the vector composed by the expected values of $w_{\rm DE}$ in each bin. To avoid any specific choice for $\mathbf{\bar{w}}$, which could affect the final results, we define this as the mean value of the EoS in the bin under consideration and in the adjacent ones, i.e.
\begin{equation}
\bar{w}_i = 
\begin{cases}
 (w_i + w_{i+1})/2 & i = 1 \\
 (w_{i-1} + w_{i} + w_{i+1})/3 & i = 2, .., N-1 \\
 (w_{i-1} + 2w_{i})/3 & i = N \\
\end{cases}
\end{equation}
where in the case of the last bin, we account for the fact that $w(a<a_N)=w(a_N)$.
Including this prior distribution, means that we are assuming that our $w_{i}$ parameters are Gaussian distributed and their fluctuations around $\mathbf{\bar{w}}$ are described by the covariance matrix $\mathcal{C}$. Since we want this matrix to encode the theoretical correlation of the values of $w_{\rm DE}(a)$ at different redshifts, $\mathcal{C}$ needs to be obtained imposing conditions arising from the physical viability of theoretical models. We use therefore the results of \citep{Articolo_priors_Ale}, where such a correlation was obtained in different Dark Energy and Modified Gravity models, imposing a set of viability and stability conditions, e.g.  requiring the avoidance of ghosts and gradient instabilities. 

Following \citep{Articolo_priors_Ale} we write the covariance matrix as
\begin{equation}
\mathcal{C}(a,a') = \sqrt{C(a)C(a')} \tilde{C}(a,a')
\end{equation}
with $C(a)$ an autocorrelation function and $\tilde{C}(a,a')$ a correlation matrix, functions only of the scale factor.

We focus our analysis on the class of canonical single field Quintessence (from now on simply Quintessence), and on the broad class of  Horndeski gravity. Despite somewhat limiting the model independence of our approach, we can still analyze a broad class of models, ensuring at the same time that the results we obtain will correspond to theoretical models which satisfy physical viability conditions.

We use $C(a)$ and $\tilde{C}(a,a')$ obtained by \citep{Articolo_priors_Ale}, where the numerical correlations found for different classes of DE and MG models were encoded in the fitting formulas shown in Table \ref{tab:matrix_tab}, where for $\tilde{C}$ two possible choices are taken into account, i.e. an exponential and a CPZ parameterization. Both these choices are built in such a way that if $\delta a=a_i-a_j$ (or $\delta \ln(a)=\ln(a_i)-\ln(a_j)$) is much higher than the correlation length $\xi$ then the correlation tends to zero. The prior that best minimizes the residuals in both the Quintessence and Horndeski cases is the exponential one \citep{Articolo_priors_Ale}. 

\begin{small}
\begin{table}[ht]
\centering
	\begin{tabular}{|c|c|c|c|c|c|}
    	\hline 
		Autocorrelation function								&					 & $\alpha$ & $\beta$ 	& $\gamma$ 	& $x$ 		\\
		\hline
		\multirow{2}{*}{$ C(x) = \alpha + \beta \exp[\gamma (x-x_{0})] $}		&	Quintessence	 & $0.03$ 	& $0.3$ 	& $6.5$ 	& $a$\\
															&	Horndeski 		 & $0.05$ 	& $0.8$ 	& $2$ 		& $\ln(a)$ 	\\
	    \hline	
		Correlation matrix								&					 & $\xi$ & $n$ 	& $x$ 	& $y$  \\
		\hline
		\multirow{2}{*}{$ \tilde{C}(x,y) =\exp \left[ \left(- \frac{\vert x-y \vert}{\xi} \right)^{n} \right] $}		&	Quintessence	 & $0.7$ 	& $1.8$ 	& $a$ 		& $a'$ \\
															&	Horndeski 		 & $0.3$ 	& $1.2$ 	& $\ln(a)$ 	& $\ln(a')$ \\ 	 
		\hline
		\multirow{2}{*}{$ \tilde{C}(x,y) = \frac{1}{1 + \left (\frac{\vert x-y \vert}{\xi} \right)^{n}}$}		&	Quintessence	 & $0.6$ 	& $2$ 	& $a$ 		& $a'$ \\
															&	Horndeski 		 & $0.2$ 	& $2$ 	& $\ln(a)$ 	& $\ln(a')$ \\
	    \hline	
	\end{tabular}
\caption{Summary of the autocorrelation and correlation analytical fits obtained by \citep{Articolo_priors_Ale}. The first parameterization of the correlation matrix is the exponential, while the second is the CPZ \citep{Crittenden_2009}. Here $x_{0}$ denotes the $x$ variable evaluated at present time, i.e. $x_{0}= a_{0} = 1$ (or $x_{0} = \ln(a_{0}) = 0$).}\label{tab:matrix_tab}
\end{table}
\end{small}

We want to stress here that having a correlation length set by theoretical requirements significantly helps in the choice of the binning strategy, since the correlation length defines a number of effective degrees of freedom $N_{\rm{eff}} = (a_{max}-a_{min})/\xi$; as long as the number of bins $N$ satisfies $N > N_{\rm{eff}}$, the dependence of the reconstruction on the number of bins is negligible \cite{Crittenden_2009}.

The specific analysis strategy will depend on the class of models under consideration. Quintessence models modify the background expansion of the Universe with respect to $\Lambda$CDM without modifying the equations for the evolution of cosmological perturbations; therefore we can use both background (SN and BAO) and perturbation (CMB) data. On the contrary, for Horndeski models the use of  CMB  data would require us to consider also their impact on the growth of cosmological structures, eventually jointly binning the phenomenological functions $\mu$ and $\Sigma$ \citep{Planck2015_DE_MG,Pogosian_Silvestri}. Since the reconstruction of these two functions is beyond the scope of this paper, we limit our analysis in this case to background data only. 

The two classes of models also have different requirements for the prior range:
\begin{itemize}
\item \textbf{Quintessence}: the EoS lies in the region above the Phantom divide ($w=-1$), hence we impose $w_{\rm DE}(a)\geq-1$. To this purpose we sample the $w_i$ parameters in the range $[-1,0]$. The reconstruction of $w_{DE}(a)$ for Quintessence is done using both the CPZ and exponential correlation priors (Table \ref{tab:matrix_tab}), to highlight how they differently affect the inferred $w_{i}$ parameters. Given the theoretical correlation length for this case ($\xi_{exp} = 0.6 \sim \xi_{CPZ}=0.7$) the $w_i$ parameters are defined in the bins $\vec{a}=(0.85,~0.7,~0.55,~0.4,~0.25,~0.1)$ in order to satisfy the requirement $N>N_{\rm{eff}}$.
\item \textbf{Horndeski}: the EoS has no particular restriction in the values that it can assume, hence $w_{i}$ is sampled in the range $[-3,0]$. The reconstruction for the Horndeski case is obtained using the exponential prior (Table \ref{tab:matrix_tab}). In this case the Planck CMB data are not included since we are not modelling the impact of this class of models on the evolution of perturbations, hence we only rely on background data, as Type Ia SNe and BAO. Concerning the latter, it is known that BAO depend on the assumption of a fiducial cosmology in order to compare data and theoretical predictions. We assume here that a change of the perturbations evolution as produced by Horndeski models does not affect significantly the use of this probe. This assumption is however deeply under scrutiny at present, and we refer the reader to recent works that investigate this problem \citep{Anselmi:2018vjz,Carter:2019ulk}. Also in this case, the choice of the scale factor range in which we define our bins is made in such a way to satisfy the requirement $N>N_{\rm{eff}}$; as the correlation length in this case is $\xi_{exp} = 0.3$ we use $\vec{a}=(0.9,~0.8,~0.7,~0.6,~0.5,~0.4)$. In this case, since we are not using Planck data, the redshift range in which data are available is smaller than in the previous case, therefore we limit our reconstruction to $a_6=0.4\ (z_{6} = 1.5) $.

\end{itemize}

We apply this prior information to our analysis both when using the smoothed step and GP reconstruction. It is worth stressing that in the latter case, the GP and prior both exploit a correlation between the values of $w_{DE}(a)$ at different scale factors. However, while GP returns the value of $w_{\rm DE}$ at a given $a_{*}$ given the $w_i$, without acting on the $w_i$ itself, the correlation prior is related to the relative values of the different $w_i$, penalizing those configurations of binned values which do not satisfy the viability conditions. In this sense, the GP and correlation priors are complementary and their combined use is possible.

\section{Results} \label{sec:results}
In this Section we present the results obtained following the analysis method described in the previous two Sections, discussing separately the reconstruction within  the Quintessence realm and that for the  more general Horndeski class.

\subsection{Quintessence}
\label{sec:Quintessence}

We report the results for Quintessence in Table \ref{tab:quint_res}. The results  are obtained  both with and without the inclusion of the theoretical prior (exponential and CPZ), for comparison. In all three cases, we find only upper limits on the $w_i$ parameters, with the lower bound $w=-1$ included within the $1-\sigma$ bound, thus indicating that these results are compatible with $\Lambda$CDM.

\begin{table}[h!]
\begin{center}
	\begin{tabular}{lccc}
	    \hline
	    \multicolumn{4}{c}{SMOOTHED STEP FUNCTION RECONSTRUCTION} \\
    	\hline 
		Parameter	&	{\small Planck+BAO+SN}	&	{\small Planck+BAO+SN (exp prior)}	&	{\small Planck+BAO+SN (CPZ prior)}	\\
		\hline
        $\Omega_{b}h^{2}$.........	&	$0.02244\pm 0.00021$ 	        &$0.02244\pm 0.00020$	   &	$0.02242\pm 0.00021$	\\
        $\Omega_{c}h^{2}$.........	&	$0.1164\pm 0.0016$	        &$0.1166\pm 0.0015$	   &	$0.1167\pm 0.0015$	\\
        $H_{0}$	...........		&	$64.9^{+1.6}_{-1.1}$	        &$65.2^{+1.6}_{-0.96}$	   &	$65.3^{+1.4}_{-0.95}$	\\
        $\Omega_{\Lambda}$............	&	$0.669^{+0.017}_{-0.011}$	&$0.671^{+0.017}_{-0.010}$ &	$0.672^{+0.015}_{-0.010}$\\
        $\Omega_{m}$............	&	$0.331^{+0.011}_{-0.017}$	&$0.329^{+0.010}_{-0.017}$ &	$0.328^{+0.010}_{-0.015}$\\
	\hline
        $w_{1}$........	                &	$<-0.745$	                &$<-0.774$	           &$<-0.790$	\\
        $w_{2}$........	                &	$<-0.921$	                &$<-0.920$	           &$<-0.918$	\\
        $w_{3}$........	                &	$<-0.911$	                &$<-0.912$	           &$<-0.915$	\\
        $w_{4}$........	                &	$<-0.888$	                &$<-0.891$	           &$<-0.901$	\\
        $w_{5}$........	                &	$<-0.806$	                &$<-0.826$	           &$<-0.844$	\\
        $w_{6}$........         	&	$<-0.700$	                &$<-0.728$	           &$<-0.751$	\\
	    \hline	    
	\end{tabular}
\caption{Mean values and $1\sigma$ confidence levels or upper limits of the Quintessence case inferred parameters, using the datasets without and with the priors, reconstructing the Equation of State via smoothed step function. The input scale factors associated to the $\vec{w}$ are $\vec{a}=(0.85,~0.7,~0.55,~0.4,~0.25,~0.1)$.}\label{tab:quint_res}
\end{center}
\end{table}

\begin{figure} [h!]
\begin{center}
\includegraphics[width=0.8\columnwidth]{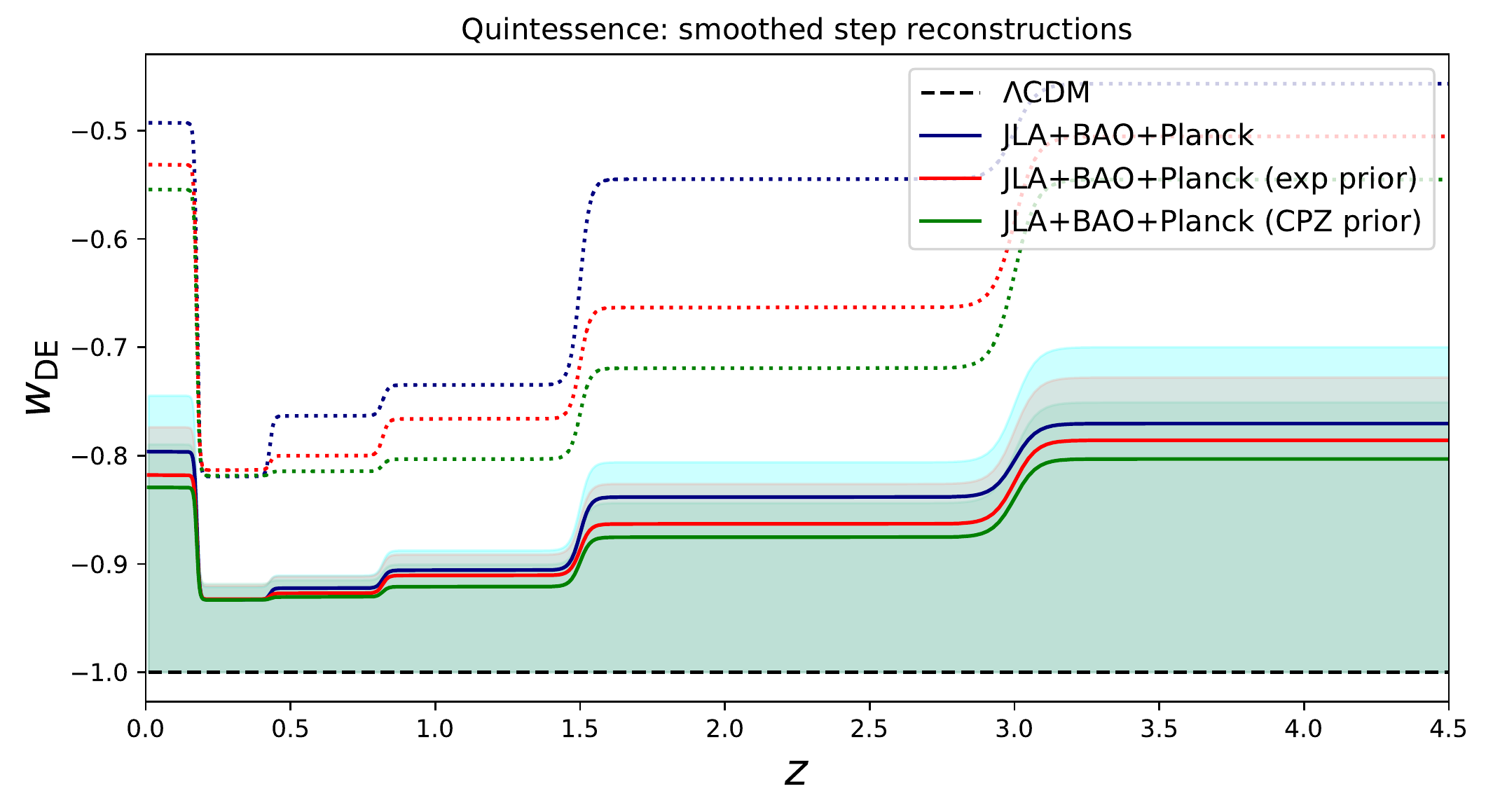}
\caption{Reconstructed mean values of $w_{\rm DE}(a)$ in the Quintessence case, obtained via smoothed step function. The blue line corresponds to the reconstruction without the inclusion of any theoretical prior on the correlation of $w_i$ parameters, while the red and green line are obtained including the exponential and CPZ correlations respectively. The filled areas of the corresponding colors trace the $1\sigma$ confidence levels, while the dotted lines delimit the $2\sigma$ confidence levels. The input scale factors are $\vec{a}=(0.85,~0.7,~0.55,~0.4,~0.25,~0.1)$.}\label{fig:quint_binned}
\end{center}
\end{figure}

Figure \ref{fig:quint_binned} shows the results obtained reconstructing the DE EoS via the smoothed step function. We point out that the mean values of the reconstruction deviate from the $\Lambda$CDM limit given the requirement of a hard bound for the posterior at $w_i=-1$, which however does not prevent $\Lambda$CDM from being a good fit for this reconstruction. 

The comparison of the three cases, bin by bin, can be quantified as a percentage difference of the mean values inferred adding the exponential (exp) and CPZ priors with respect to the mean values obtained with no theoretical prior. The main differences associated to the choice of a particular prior are related to the behavior of $w_{4}$ (exp: $-1.32\%$, CPZ: $-0.44\%$), $w_{5}$ (exp: $-3.3\%$, CPZ :$-2.01\%$) and $w_{6}$ (exp: $-4.3\%$, CPZ: $-2.47\%$) mean values and $1\sigma$ confidence levels, while $w_{1}$ (exp: $-0.249\%$, CPZ: $0\%$), $w_{2}$ (exp: $+0.214\%$, CPZ: $+0.107\%$) and $w_{3}$ (exp: $0\%$, CPZ: $+0.108\%$) do not differ significantly case by case.
The larger impact of the correlation priors in the last three bins can be explained with the fact that these redshifts contain much less data; therefore the constraining power of the prior is comparable to that of the data.

The prior however, does not impact significantly the behavior of the function, showing how the requirement $w_i\geq-1$ already satisfies the physical viability conditions imposed through the correlation prior.

In order to test the dependence of the results on the reconstruction method, we perform the analysis for the Quintessence class, both with Planck+BAO+SN only and with the inclusion of the CPZ prior, using the Gaussian Process reconstruction method rather then the smoothed bins one. In this case, we fix the correlation length $\xi$ of the GP to that of the theoretical prior. Furthermore, we choose to force $w_{\rm DE}(a)$ to remain constant at the value of the first redshift bin $w_1$ for $a>a_1$.

\begin{table}[ht]
\centering
	\begin{tabular}{lcc}
	    \hline
	    \multicolumn{3}{c}{GAUSSIAN PROCESS RECONSTRUCTION} \\
    	\hline 
	Parameter	                &{\small Planck+BAO+SN}	         &{\small Planck+BAO+SN (CPZ prior)}	\\
	\hline
        $\Omega_{b}h^{2}$.........	&$0.02242^{+0.00022}_{-0.00020}$ &$0.02241\pm 0.00021$	\\
        $\Omega_{c}h^{2}$.........	&$0.1166\pm 0.0015$	         &$0.1167\pm 0.0014$		\\
        $H_{0}$	...........		&$62.8^{+1.6}_{-2.0}$		 &$63.4^{+2.2}_{-1.6}$			\\
        $\Omega_{\Lambda}$............	&$0.645\pm 0.020$		 &$0.651^{+0.025}_{-0.017}$	\\
        $\Omega_{m}$............	&$0.355\pm 0.020$		 &$0.349^{+0.017}_{-0.025}$ 	\\
	\hline
        $w_{1}$........	                &$<-0.86$                        &$<-0.89$		\\
        $w_{2}$........          	&$<-0.91$                        &$<-0.91$		\\
        $w_{3}$........	                &$<-0.91$	                 &$<-0.91$		\\
        $w_{4}$........	                &$<-0.91$	                 &$<-0.91$		\\
        $w_{5}$........	                &$<-0.81$	                 &$<-0.84$		\\
        $w_{6}$........	                &$<-0.63$ 	                 &$<-0.73$		\\
	\hline	    
	\end{tabular}
\caption{Mean values and $1\sigma$ confidence levels of the Quintessence case inferred parameters, using the datasets without and with the CPZ prior, reconstructing the Equation of State via Gaussian Process method. The input scale factors associated to the $\vec{w}$ are $\vec{a}=(0.85,~0.7,~0.55,~0.4,~0.25,~0.1)$.}\label{tab:results_GP_Quint}
\end{table}

\begin{figure} [h!]
\begin{center}
\includegraphics[width=0.85\textwidth]{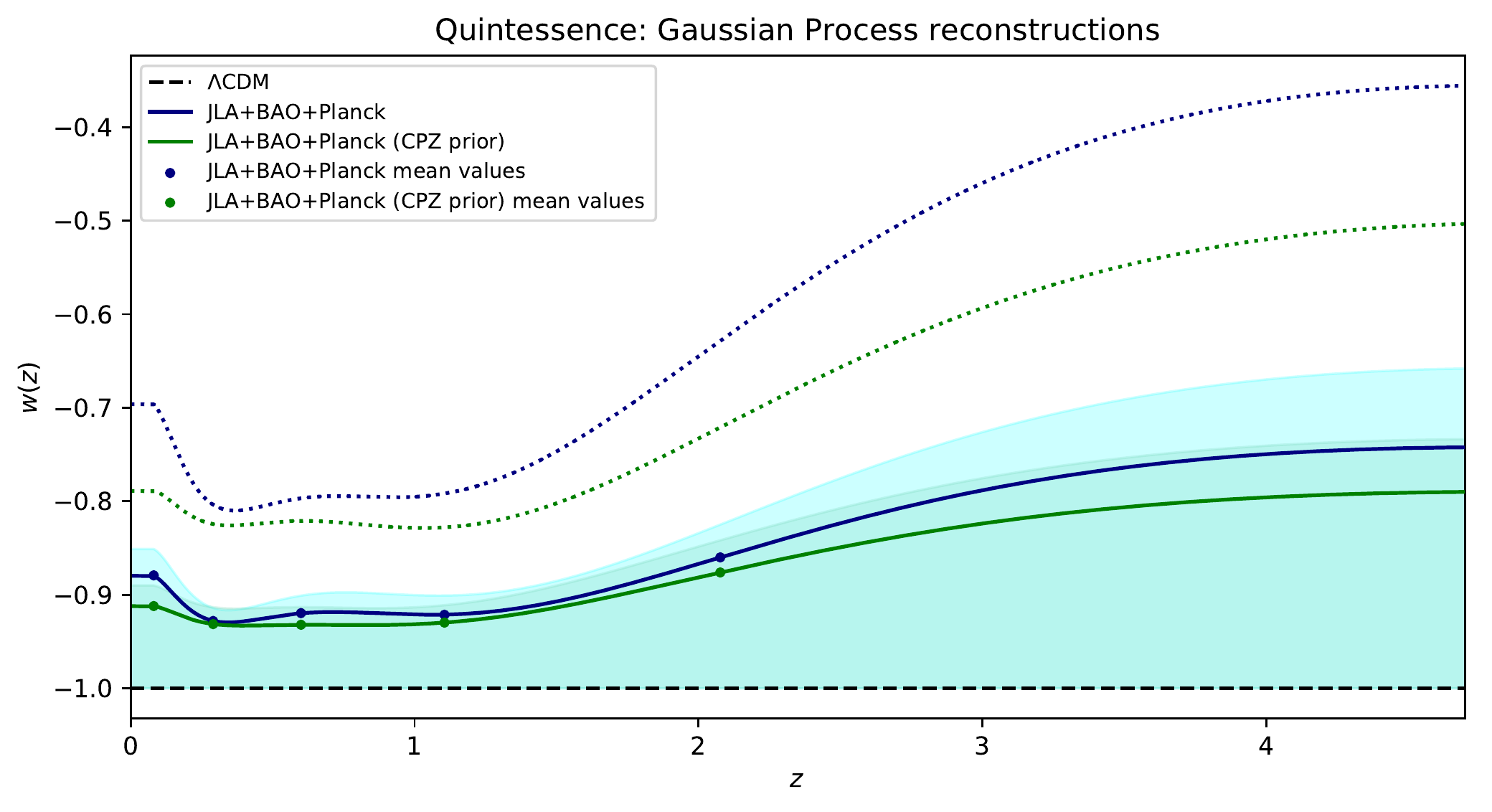}
\caption{Reconstructed mean values of $w_{\rm DE}(a)$ in the Quintessence case, obtained via Gaussian Process. The cyan line corresponds to the reconstruction without the inclusion of any theoretical prior on the correlation of $w_i$ parameters, while the lime line is obtained including the CPZ correlation. The filled areas of the corresponding colors trace the $1\sigma$ confidence levels, while the dotted lines delimit the $2\sigma$ confidence levels. The input scale factors are $\vec{a}=(0.85,~0.7,~0.55,~0.4,~0.25,~0.1)$.}\label{fig:Quint_GP_single}
\end{center}
\end{figure}

\begin{figure} [!h]
\begin{center}
\includegraphics[width=0.85\textwidth]{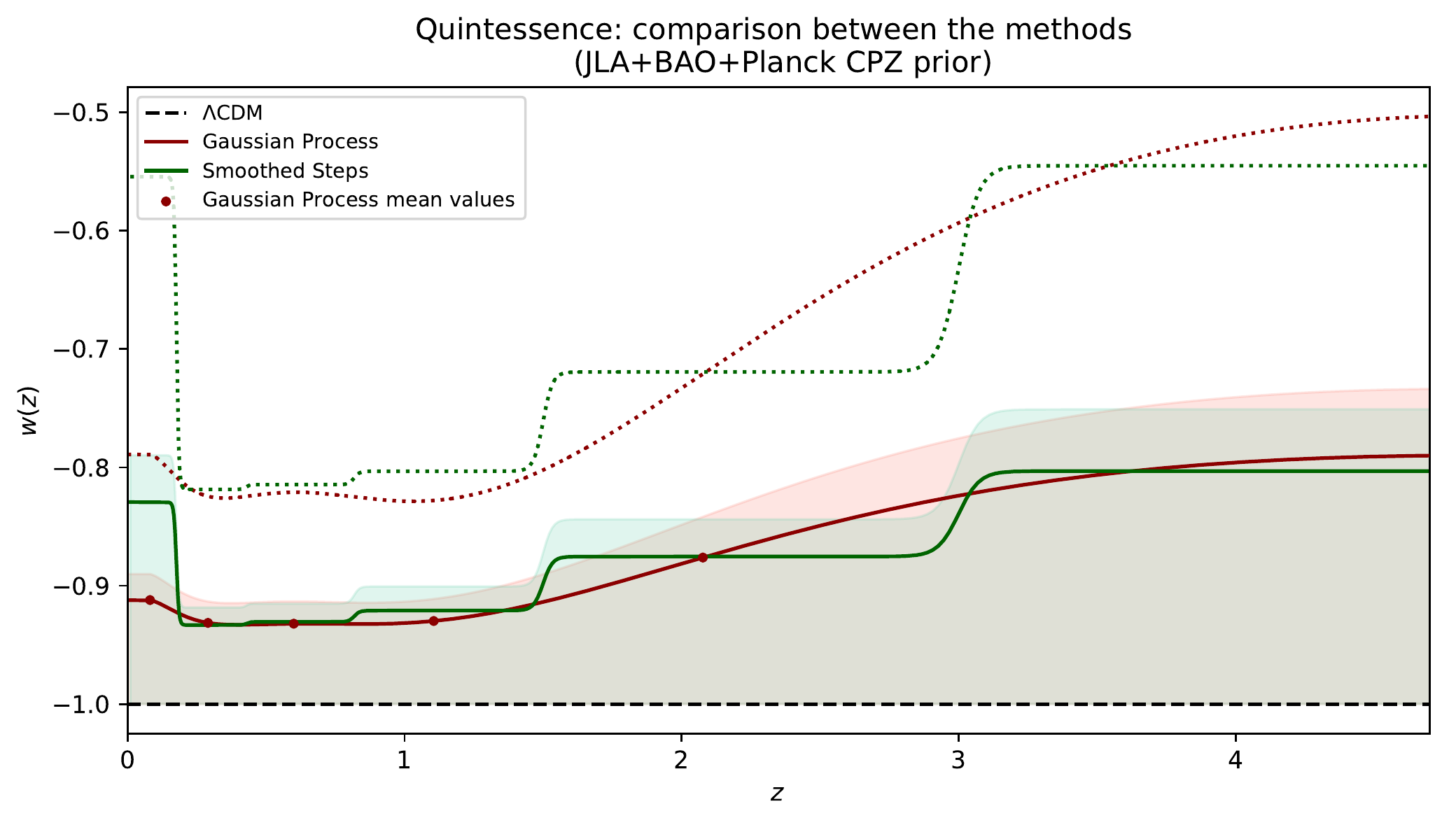}
\caption{Superposition of the smoothed step function and Gaussian Process reconstructions for Quintessence Equation of State obtained including the datasets and the CPZ prior. For each case, the continuous lines are the reconstructions obtained using the mean values, whereas the filled areas of the corresponding colours trace the $1\sigma$ confidence levels, while the dotted lines delimit the $2\sigma$ confidence levels. The input scale factors are $\vec{a}=(0.85,~0.7,~0.55,~0.4,~0.25,~0.1)$.}\label{fig:Quint_cfr_GP_smoothed}
\end{center}
\end{figure}

Table \ref{tab:results_GP_Quint} shows the results obtained on the cosmological and reconstruction parameters in the Planck+BAO+SN and Planck+BAO+SN+prior cases, while in Figure \ref{fig:Quint_GP_single} we show the reconstruction of the $w_{\rm DE}(a)$ function given the mean values of $w_i$. We notice how the results are compatible with the smoothed bins reconstruction, but with the prior affecting more significantly than before the earliest redshifts of the reconstruction: the mean values of the reconstruction obtained adding the prior respectively differ from the $w_{1}$, $w_{2}$ and $w_{3}$ means obtained with only the datasets by $+4.2\%$, $-1.5\%$ and $+0.86\%$. These results are emphasized by the comparison shown in Figure \ref{fig:Quint_cfr_GP_smoothed}.

\begin{figure} [!h]
\begin{center}
\includegraphics[width=0.85\textwidth]{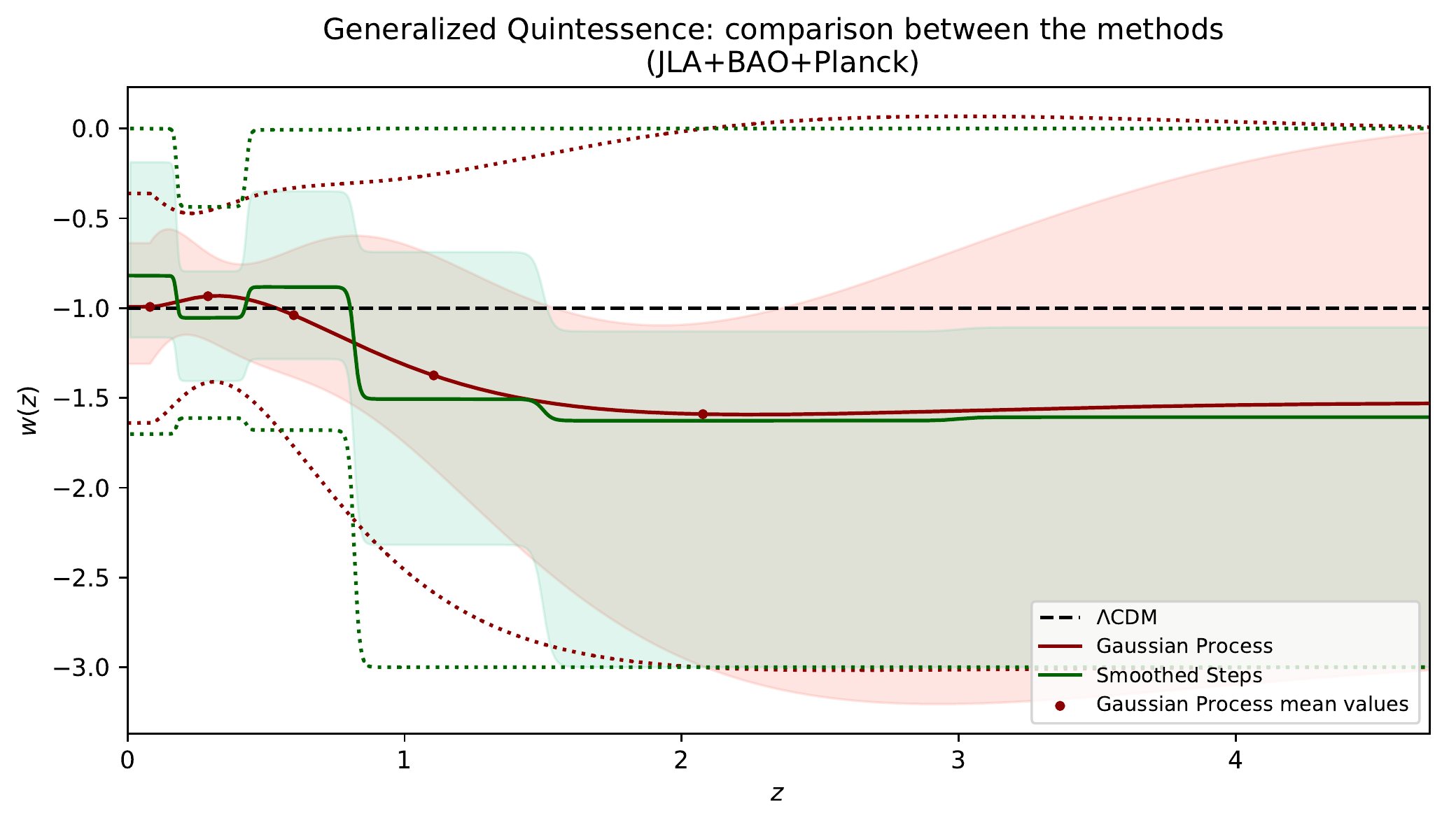}
\caption{Superposition of the smoothed step function and Gaussian Process reconstructions for the {\it Generalized Quintessence} case using the Planck+BAO+SN datasets combination. For each case, the continuous lines are the reconstructions obtained using the mean values, whereas the filled areas of the corresponding colours trace the $1\sigma$ confidence levels, while the dotted lines delimit the $2\sigma$ confidence levels. The input scale factors are $\vec{a}=(0.85,~0.7,~0.55,~0.4,~0.25,~0.1)$.}\label{fig:genquint}
\end{center}
\end{figure}

In addition to these results obtained in the single field minimally coupled Quintessence case, a model implying a limitation $w_i\geq-1$, we show in Figure \ref{fig:genquint} the reconstruction obtained with the Planck+BAO+SN dataset when this requirement is not imposed, labelling it {\it Generalized Quintessence}. EoS crossing the phantom divide can be obtained by models with non minimal coupling or with kinetical braiding \citep{Carroll:2004hc,Deffayet:2010qz,Easson:2016klq}, as well as by scalar-tensor gravity \cite{Boisseau:2000pr}. Notice however that in general these models might introduce also modifications of perturbation evolution with respect to $\Lambda$CDM, which are not included in our reconstruction. We stress therefore that the results presented here assume that even with such a generalized expansion history, no modification of perturbations is produced. Moreover, as we have no available theoretical information on this reconstruction case, we do not make use of theoretical priors. We find that also in this case the two reconstruction methods are compatible with each other, obtaining the same trend in redshift for the EoS within the $1\sigma$ bound. The $\Lambda$CDM limit $w=-1$ is generally compatible with the results found within $1\sigma$ except for the bin at $a=0.25$, where the EoS is constrained to be below $w=-1$ at $1\sigma$. Nevertheless, no evidence for deviation from standard cosmological constant is found, as $\Lambda$CDM is always compatible within the $2\sigma$ limit.

\subsection{Horndeski class of MG models}
\label{sec:Horn}

In Table \ref{tab:results_Horn} we report the results obtained reconstructing the EoS via smoothed step function and GP in the Horndeski case, for which the $w_{i}$ parameters are free to vary within the range $[-3,0]$ and the exponential correlation prior is imposed. As already mentioned, in this case we restrict to background data, and do not use CMB measurements.
In both cases the last bin considered $w_{6}$ is unconstrained due to the fact that very few data are available at this redshift ($z \sim [1,1.5]$). Because of the correlation that we have introduced between the binned values of $w_{\rm DE}$, this lack of constraining power propagates also to the constraints at higher scale factors (lower redshifts).
For the other bins, we obtain compatible constraints from the two reconstruction methods, in agreement within $1\sigma$ with each other and with a $\Lambda$CDM expansion history. Notice that in this case, with $w_{\rm DE}(a)$ able to take values below $-1$ and with no CMB data, we do not have constraining power on $H_0$.

\begin{table}[!h]
\begin{center}
	\begin{tabular}{lcc}
    	\hline 
        \multicolumn{3}{c}{RECONSTRUCTIONS: JLA+BAO (exp prior)} \\
        \hline
		Parameter	&	\small{Smoothed step function}	&	\small{Gaussian Process} 	\\
		\hline
        $\Omega_{b}h^{2}$.........	&	$0.041^{+0.012}_{-0.031}$ 	& $0.049^{+0.017}_{-0.031}$	\\
        $\Omega_{c}h^{2}$.........	&	$0.181^{+0.084}_{-0.12}$ 	& $0.157^{+0.067}_{-0.10}$	\\
        $H_{0}$	...........			&	$> 76.6$		& $> 79.7$	\\
        $\Omega_{\Lambda}$............	&	$0.684\pm 0.086$		& $0.713\pm 0.072$	\\
        $\Omega_{m}$............		&	$0.316\pm 0.086$	& $0.287\pm 0.072$	\\
		\hline
        $w_{1}$........	&	$-1.10^{+0.30}_{-0.22}$				& $-1.10^{+0.28}_{-0.23}$	\\
        $w_{2}$........	&	$-0.97^{+0.36}_{-0.22}$				& $-0.86^{+0.26}_{-0.18}$	\\
        $w_{3}$........	&	$-1.09^{+0.60}_{-0.30}$				& $-0.98^{+0.44}_{-0.25}$	\\
        $w_{4}$........	&	$-1.21^{+0.73}_{-0.39}$				& $-1.05^{+0.58}_{-0.34}$	\\
        $w_{5}$........	&	$-1.44^{+0.75}_{-0.63}$				& $-1.36^{+0.74}_{-0.62}$	\\
        $w_{6}$........	&	unconstrained					& unconstrained	\\
	    \hline	    
	\end{tabular}
\caption{Mean values and $1\sigma$ confidence levels of the Horndeski case, reconstructing the Equation of State via smoothed step function and via Gaussian Process. The input scale factors associated to the $\vec{w}$ are $\vec{a}=(0.9,~0.8,~0.7,~0.6,~0.5,~0.4)$. Here $w_{6}$ is unconstrained, with the $1\sigma$ region extending over the full prior range.}\label{tab:results_Horn}
\end{center}
\end{table}

\begin{figure} [!h]
\begin{center}
\includegraphics[width=0.75\textwidth]{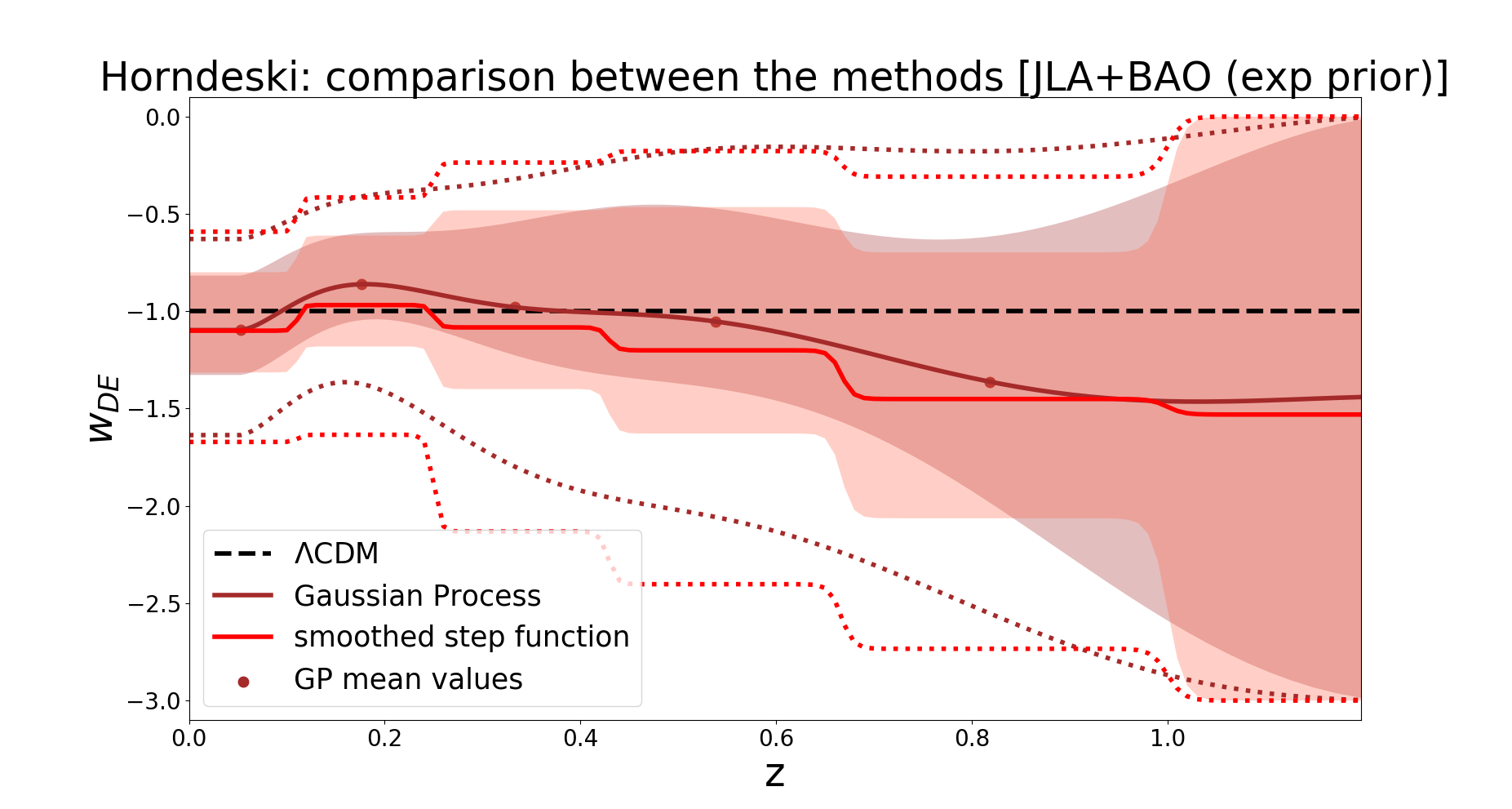}
\caption{Superposition of the smoothed step function and Gaussian Process reconstructions for the Equation of State in the Horndeski case, obtained including the datasets and the exponential prior. For each case, the continuous lines are the reconstructions obtained using the mean values, whereas the filled areas of the corresponding colours trace the $1\sigma$ confidence levels, while the dotted lines delimit the $2\sigma$ confidence levels. The input scale factors are $\vec{a}=(0.9,~0.8,~0.7,~0.6,~0.5,~0.4)$.}\label{fig:Horn_cfr_gpsmoo}
\end{center}
\end{figure}

In Figure \ref{fig:Horn_cfr_gpsmoo} we show the reconstruction of $w_{DE}(a)$ given the inferred $w_{i}$ using both the smoothed and Gaussian process reconstruction methods. The function tends to decrease over redshift, independently of the reconstruction method, with the mean reconstructed function slightly closer to the $\Lambda$CDM value in the GP case for intermediate redshifts. We stress however that both reconstruction are compatible between each other and with the $\Lambda$CDM limit at approximately $1\sigma$. 

The inclusion of CMB data also in the Horndeski case would increase the amount of information and potentially improve the constraints achievable with this method; however, as already pointed out, this would require the reconstruction of the $\mu(a)$ and $\Sigma(a)$ encoding the departures from the standard $\Lambda$CDM growth of cosmological perturbations. The possible degeneracies between the effects of these two functions and those of a varying $w_{\rm DE}(a)$ can therefore limit the improvement in the achievable constraints when CMB data are included. A full analysis of this scenario would therefore require the use of cosmological data able to disentangle the background and perturbation modifications, e.g. Cosmic Shear or Galaxy Clustering data from Large Scale Structure surveys \citep{Abbott_1,KiDS_1,KiDS-450}.

\newpage
\section{Conclusions} \label{sec:concl}

We have reconstructed the Equation of State (EoS) of Dark Energy,  $w_{\rm DE}(a)$, from the latest cosmological data with  two different techniques: one in which the EoS is assumed constant within the range of each specified bin in scale factor, with smooth transitions between binned values; and one in which the function is reconstructed at each value of $a$ from its binned values using a Gaussian Process reconstruction. 
We modified the public code {\tt CAMB} to produce predictions on cosmological observables given the reconstructed $w_{\rm DE}(a)$ and we compared these predictions with currently available data both for background (SN and BAO) and for perturbations (CMB) observables. Alongside the observational data, we took into account the contribution of theoretical conditions on the physical viability of the reconstructed EoS; this was possible with the inclusion of a correlation prior between the binned values of $w_{\rm DE}(a)$, obtained in two classes of models: single field Quintessence and Horndeski gravity \citep{Articolo_priors_Ale}.

The analysis of these two classes differs both in the binning strategy for $w_{\rm DE}(a)$, which is defined by the different correlation lengths $\xi$, and in the observables used to constrain the parameters: in the Quintessence case, the reconstruction of $w_{\rm DE}(a)$ was sufficient to fully characterize the modification on background and perturbations evolutions with respect to $\Lambda$CDM and we can therefore use all the data from BAO, SN and CMB. In the Horndeski case instead, the description of perturbations would have required also to reconstruct the $\mu(a)$ and $\Sigma(a)$, and therefore we limited our analysis to the background data coming from BAO and SN observations.
Moreover, while for Horndeski the binned values $w_i$ were allowed to vary in the full prior range $[-3,0]$, this range was limited in Quintessence to $[-1,0]$ as these models cannot cross the phantom divide $w=-1$ without developing ghost instabilities.

In our results for the Quintessence case, we found that $\Lambda$CDM $w=-1$ is compatible with the reconstructed $w_{\rm DE}(a)$ within $1\sigma$. Moreover we found that the theoretical prior is not affecting the results significantly, due to the limitation of our analysis to the non-phantom part of the parameter space ($w_i\geq-1$) which already satisfies the most restrictive of the physical viability conditions 
\cite{Peirone:2017lgi}. The results of the two reconstruction methods are in agreement with each other, highlighting how for a high enough number of bins, the smoothed bins reconstruction is able to reproduce sufficient variations of $w$ in redshift to fit cosmological data.

In the Horndeski case, we also found that $w=-1$ is compatible with the reconstructed EoS within $1\sigma$. The smoothed bins and GP reconstruction methods are found to be compatible also in this case,  although less than in the Quintessence analysis; this is possibly due to the fact that the oscillatory behavior hinted by the reconstruction could require a higher number of bins to be reproduced in the smoothed bins case, or that a different GP Kernel would be more appropriate for such behavior. Tighter constraints on the EoS, that would  allow to quantify the impact of the theoretical prior also for this class of models, could be obtained including CMB data in the analysis, which requires to simultaneously reconstruct the phenomenological functions describing deviations from $\Lambda$CDM in clustering and lensing,  respectively $\mu$ and $\Sigma$. As we discussed, given the degeneracies between the effects of $w$, $\mu$ and $\Sigma$, additional observables would be needed for this kind of analysis, e.g. coming from observations of Large Scale Structures.

\acknowledgments
We want to thank Simone Peirone, Marco Raveri and Robert Crittenden for useful discussions. MM acknowledges support from the D-ITP consortium, a program of the NWO that is funded by the OCW. AS acknowledges support from the NWO and the Dutch Ministry of Education, Culture and Science (OCW), and from the D-ITP consortium, a program of the NWO that is funded by the OCW.

\bibliographystyle{unsrt}
\bibliography{biblio}

\end{document}